\documentclass[aps,prd,oneside,twocolumn,superscriptaddress,floatfix,nofootinbib,showpacs]{revtex4-2}
\usepackage{amsmath}

\usepackage{bm}
\usepackage{xcolor}
\usepackage{hyperref}
\usepackage{url}
\usepackage{graphicx}
\usepackage[large]{subfigure}
\usepackage{amssymb}
\usepackage[amssymb]{SIunits}
\usepackage{natbib}
\usepackage{multirow}
\usepackage{amsmath}
 
\usepackage{cleveref}
\usepackage{simpler-wick}
\usepackage{verbatim}
\usepackage{color}
\usepackage{xcolor}
\usepackage{amsmath}
\usepackage{bm}
\usepackage{hyperref}
\usepackage{url}
\usepackage{graphicx}
\usepackage[large]{subfigure}
\usepackage{amssymb}
\usepackage[amssymb]{SIunits}
\usepackage{natbib}
\usepackage{multirow}
\usepackage{amsmath}
\usepackage{cleveref}
\usepackage{simpler-wick}
\usepackage{booktabs} 

\newcommand{\axioncamb}{\texttt{axionCAMB}}

\newcommand{\emcee}[1]{\texttt{emcee} #1}

\newcommand{\mnras}{Mon. Not. R. Astron. Soc.}
\newcommand{\jcap}{J. Cosmol. Astropart. Phys.}
\newcommand\aj{{\rm AJ}}
\renewcommand\prd{{\rm PhRvD}}
\renewcommand\prl{{\rm PhRvL}}

\newcommand{\aap}{Astron. Astrophys.}

\newcommand\pasp{{\rm PASP}}

\begin{document}

\title{Improved constraints on ultralight axions using latest observations of the early and late Universe}

\author{Qianshuo Liu}
\affiliation{Department of Astronomy,
University of Science and Technology of China, Hefei 230026, China}
\affiliation{School of Astronomy and Space Science, University of Science and Technology of China, Hefei 230026, China}

\author{Chang Feng}
\altaffiliation{Corresponding author: changfeng@ustc.edu.cn}
\affiliation{Department of Astronomy,
University of Science and Technology of China, Hefei 230026, China}
\affiliation{School of Astronomy and Space Science, University of Science and Technology of China, Hefei 230026, China}

\author{Filipe B. Abdalla}
\affiliation{Department of Astronomy,
University of Science and Technology of China, Hefei 230026, China}
\affiliation{School of Astronomy and Space Science, University of Science and Technology of China, Hefei 230026, China}

\begin{abstract}

Ultralight axions (ULAs) are hypothetical particles which can behave like dark matter (DM) or dark energy (DE) depending on masses generated at the symmetry-breaking scale. It remains a mystery whether the ULAs can make up a fraction of DM or DE. Although theoretical predictions indicate that the ULAs may leave distinct imprints on cosmological signals, these signatures may exist in a broad spatial and temporal scales, and may be degenerate with the known effects of the standard model. The ULA signatures are extremely subtle and the observational evidence of the ULAs remain elusive. In this work, we infer the ULA properties using both the early and late universe observations from the cosmic microwave background (CMB) and baryon acoustic oscillations (BAO). We validate modeling of the ULA effects using the CMB and BAO mock data and perform different tests to cross-check the results. By analyzing the Planck 2018 CMB measurements and the BAO measurements from the Data Release 2 of Dark Energy Spectroscopic Instrument (DESI), we constrain the energy density fraction ratio of the ULAs to total dark matter $\Omega_a/\Omega_d$ and obtain a new upper bound of $\Omega_a/\Omega_d$. Future CMB and BAO measurements will achieve unprecedented precision and will be crucial for understanding the nature of the ULAs.
\end{abstract}

\maketitle

\section{Introduction}
Axions were initially introduced to solve the strong charge-parity problem~\citep{strongcp} and may have profound cosmological and astrophysical implications at ultralow mass regime, where extremely long de Broglie wavelengths of the ultralight axions (ULAs) can leave unique imprints on cosmological and astrophysical observations, such as the cosmic microwave background (CMB), baryon acoustic oscillations (BAO) and spatial matter fluctuations.

The ULA can be described by a scalar field $\phi$ in a periodic potential $V(\phi)=m_a^2f^2_a[1-\cos(\phi/f_a)]$ where the ULA mass is $m_a$ and the symmetry-breaking scale is $f_a$~\citep{ulatheoryreview}. The scalar field can roll down the potential and start oscillating when it reaches the minimum, manifesting a dynamical nature. The evolution timescale of the scalar field is determined by its mass, and the scalar field with a mass of $10^{-27}$ eV can begin oscillations around the radiation-matter equality~\citep{ulatheory}.

For masses larger than $10^{-27}$ eV, the ULAs begin oscillations in the very early universe and behave like dark matter (DM). The ULA can modify the epoch of matter-radiation equality and affect the CMB acoustic peaks. Also, the DM-like ULAs can leave unique suppression features on the matter fluctuations~\citep{ulatheory,ulatheory2,renee2015}.

For masses lower than $10^{-27}$ eV, the ULAs begin oscillations in the late Universe so behave like dark energy (DE) accelerating the expansion of our universe. The DE-like ULAs can change the distance to the last scattering surface, the decaying gravitational potential for the integrated Sachs-Wolfe (ISW) effect, and the structure growth in the late universe. Thus, the locations of the CMB acoustic peaks can be shifted, the smearing effects of gravitational lensing can be adjusted, and the large-scale fluctuations can be altered. Also, the matter fluctuations may experience a global decrease due to an insufficient amount of growth time~\citep{ulatheory2, renee2015}.

These ULA effects have been searched for and the energy density fraction ratio of the ULA ($\Omega_a$) to total dark matter ($\Omega_d$), i.e., $f_{ax}=\Omega_a/\Omega_d$ has been constrained by different observational datasets. A constraint of joint CMB measurements from Wilkinson Microwave Anisotropy Probe, Planck satellite, Atacama Cosmology Telescope and South Pole Telescope, and galaxy power spectrum measurements from the WiggleZ dark energy survey~\citep{wigglez} is found to be $f_{ax}<0.048$ in mass range $10^{-33}{\rm eV}<m_a<10^{-25.5}{\rm eV}$~\citep{renee2015}. The result is further improved to $f_{ax}<0.019$ at $m_a=10^{-28}$ eV with Planck 2018 data release~\citep{renee2018}. By adding monopole and quadrupole moments of galaxy clustering measurements, the ULA fraction constraint is further pushed down to $f_{ax}<0.014$ at $m_a=10^{-28}$ eV~\citep{p0p2}. With better sensitivities coming from Planck 2018 CMB data release and higher-order galaxy power spectra from Baryon Oscillation Spectroscopic Survey (BOSS)~\citep{boss}, an improved constraint is found to be $f_{ax}<0.008$ at $m_a=10^{-28}$ eV~\citep{rogersetal2023}. In addition, small-scale Lyman-$\alpha$ measurements yield a new limit $f_{ax}\sim 0.01\mbox{-}0.05$ at a much higher mass $m_a=10^{-25}$ eV~\citep{lyaconstraint}. Different from the fluctuations and distance measurements, a recent analysis shows that cluster number counts can also place an interesting limit $f_{ax}<0.013$ at $m_a=10^{-28}$ eV~\citep{clusterconstraint}.

The ULA dark energy nature implies that it could be an evolving dark energy component. It is unclear if the recent evidence of dynamical dark energy (DDE) from the DESI data release 2 (DR2) can be partially accounted for by the ULA species~\citep{desi2}. In this work, we run a Bayesian analysis to infer the ULA fraction $f_{ax}$ from the Planck 2018 CMB and DESI-DR2 datasets and infer the DDE properties in the presence of the ULA component.

\section{Data sets and method validations}
\label{Method}

There are multiple numerical implementations of the ULA physics, such as \texttt{axionCAMB}~\citep{renee2015} and \texttt{AxiCLASS}~\citep{axiclass}. In this work, we adopt \texttt{axionCAMB} as the baseline method to derive statistical properties of ULA physics. Effective fluid approximations adopted in the \texttt{axionCAMB} have been recently validated by exact solutions of the ULA fields~\citep{whu2025axion}.

We can describe the ULA evolution with a cosine-like potential $V(\phi)=m_a^2f_a^2[1-\cos{(\phi/f_a)}]^n$ with $n=1$, which can be approximated as $V(\phi)\sim (1/2)m_a^2\phi^2$. The ULAs with potential $n>1$ will become a radiation relic today~\citep{axiclass}. The CMB power spectra with ULA physics can be calculated with the dynamical evolution of the ULA field included in the Boltzmann equation. We use \texttt{axionCAMB} to compute lensed power spectra for CMB observables, time-dependent Hubble parameter and comoving distance for the BAO observables. 

\subsection{CMB and BAO measurements}

We use Planck 2018 CMB power spectra which contain $C_{\ell}^{TT}$ within the multipole range $30\le\ell\le2508$, and $C_{\ell}^{EE}$ and $C_{\ell}^{TE}$ within the multipole range $30\le \ell\le 1996$~\citep{pk2018}. In addition, we include the large-scale power spectra $C_{\ell}^{TT}$ and $C_{\ell}^{EE}$ at $2\le\ell\le 29$. We do not include the Planck lensing measurement since it can only negligibly affect the limits on the ULA fractions. We use the Planck released likelihood function $\texttt{Plik}$ to compute $\mathcal{L}_{\rm CMB}(\hat d_{\rm CMB}|\theta)$~\citep{pk2018}. Here $\hat d_{\rm CMB}$ refers to the measured CMB power spectra and $\theta$ denotes the cosmological parameters.

\begin{figure}
\includegraphics[width=8cm,height=8cm]{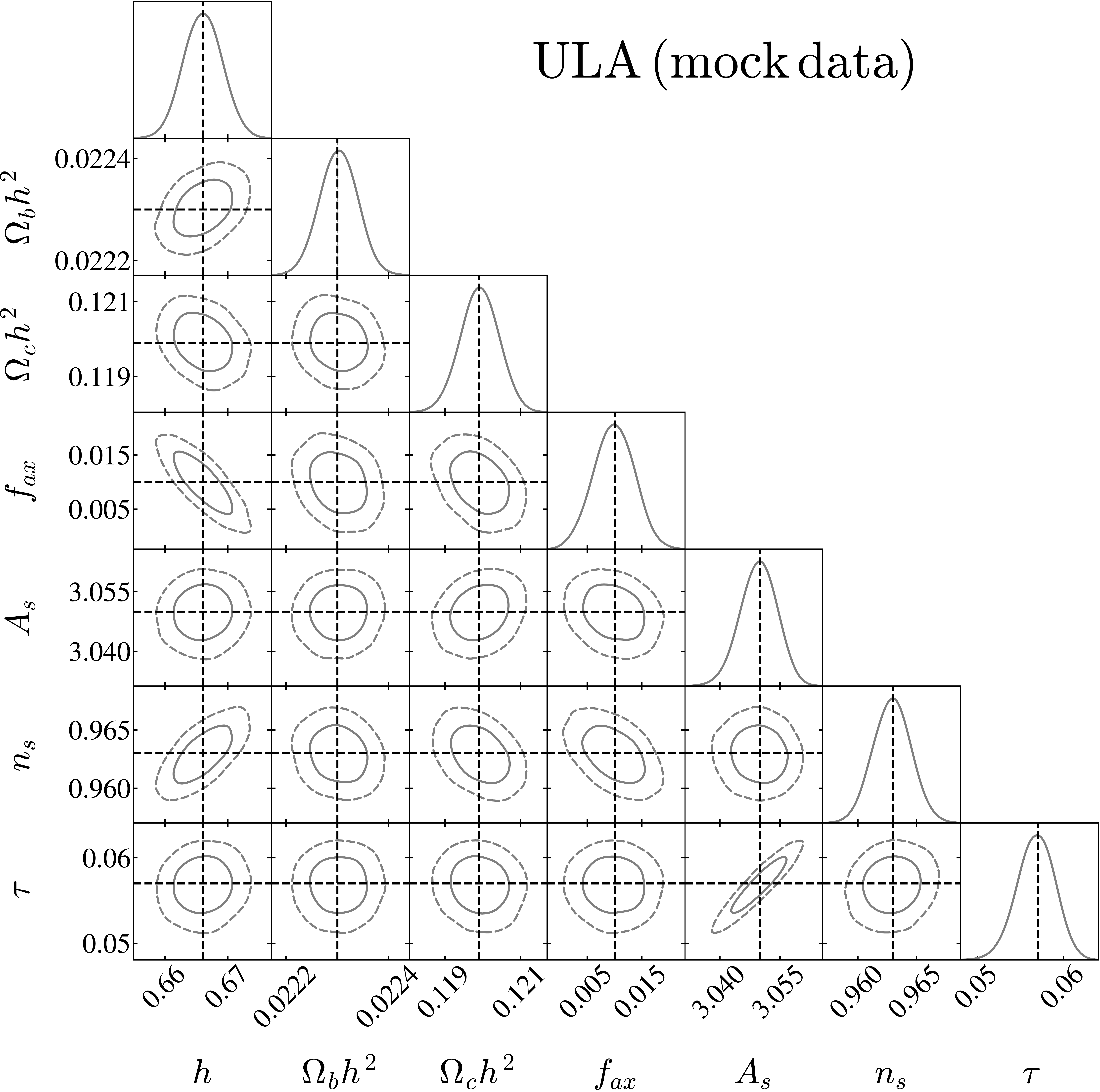}
\includegraphics[width=8cm,height=8cm]{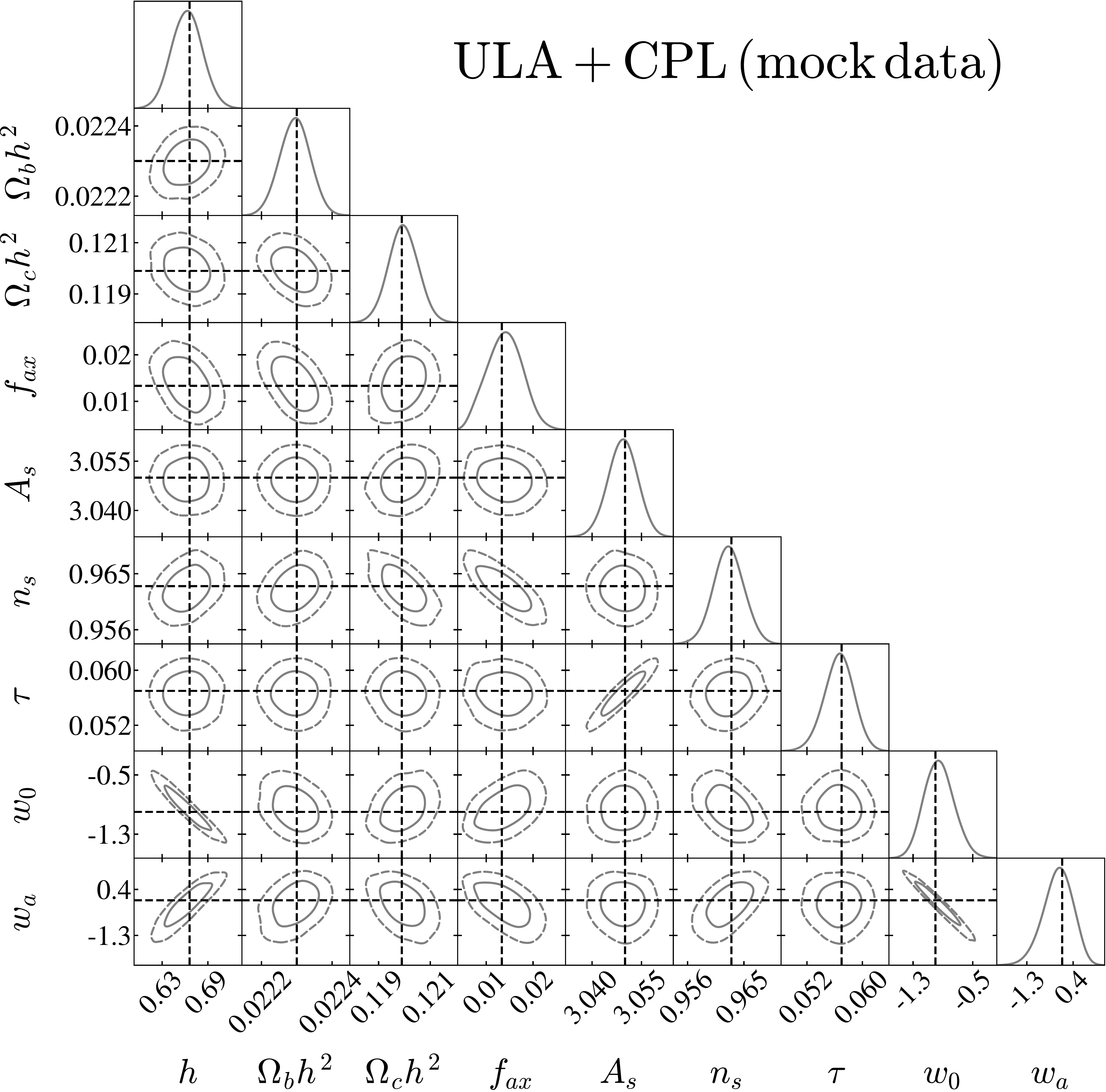}
\caption{Parameter inference from CMB and BAO mock datasets. In this test, we generate the mock data assuming an ULA fraction $\Omega_a/\Omega_d=0.01$ at a representative mass $m_a=10^{-28}$ eV. The inferred contours are consistent with the fiducial values, verifying that the Markov chain Monte Carlo (MCMC) analysis can fully recover the input cosmological parameters for the ULA masses considered in this work.} 
\label{validation}
\end{figure}

\begin{figure*}
\includegraphics[width=8cm,height=7cm]{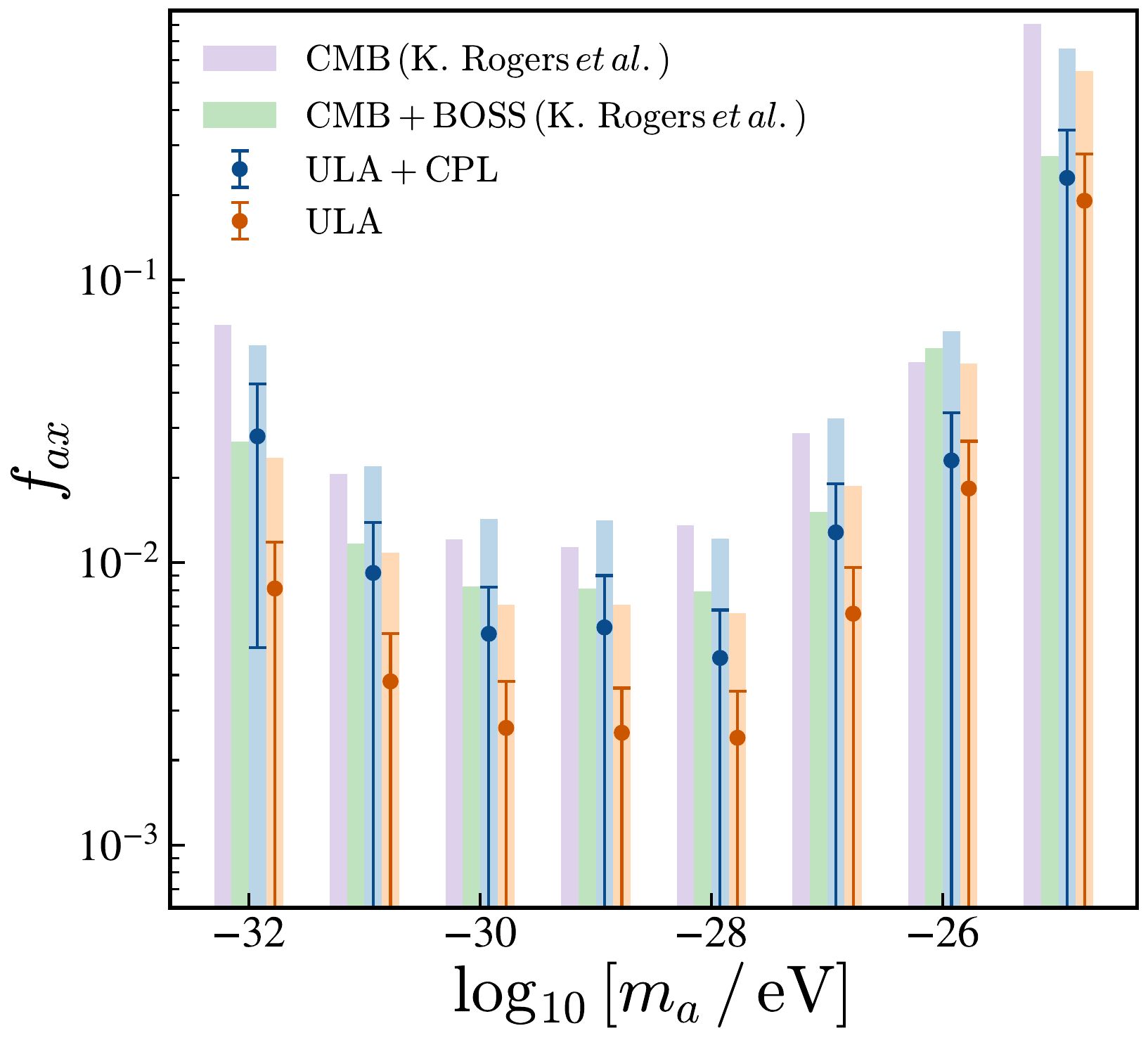}
\includegraphics[width=8cm,height=7cm]{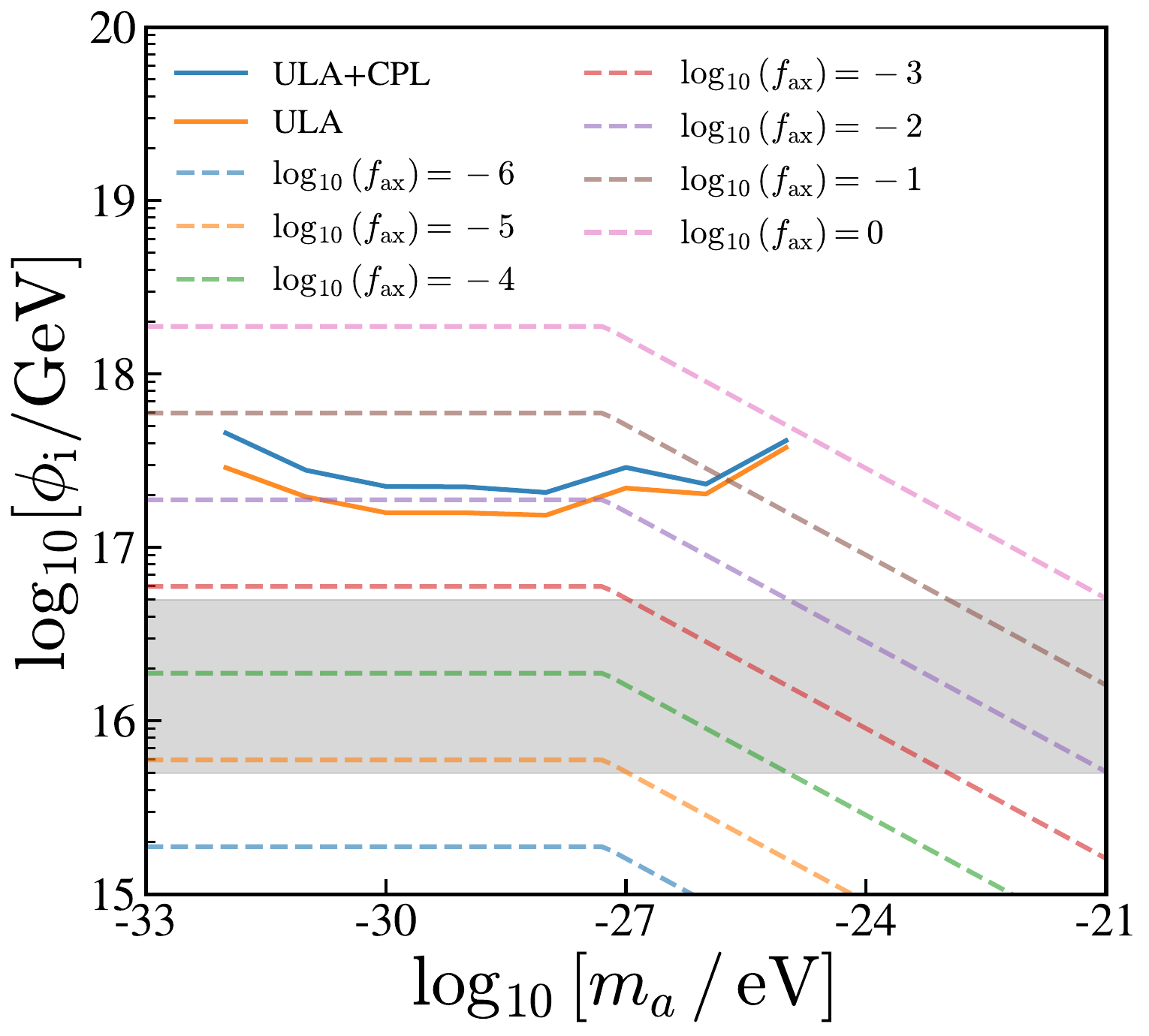}
\caption{(Left) 95\% C.L. upper bounds on the ultralight axion (ULA) energy density fraction \( f_{ax} = \Omega_a/\Omega_d \) across a range of masses. Results are shown for both \texttt{ULA}-only (orange) and \texttt{ULA+CPL} (blue) scenarios. The colored bands refer to the 95\% upper bounds and the data points with error bars denote the best-fit values with 1$\sigma$ uncertainties. The upper bounds from Planck+BOSS~\citep{rogersetal2023} are also shown for comparison. (Right) The inferred symmetry-breaking scale from the ULA energy density fractions. We use the $\Omega_a\mbox{-}\phi_i$ relation derived in~\citep{renee2015} to estimate the initial values of the ULA field $\phi_i$ which is related to the symmetry-break scale $f_a$ by $\phi_i=\theta_i f_a$ with the initial misalignment angle $\theta_i$ being order of unity. The shaded region is the grand unified theory (GUT) energy scale between $10^{15.5}<f_a<10^{16.5}$ GeV. }
\label{fractionplot}
\end{figure*}

\begin{figure*}
\includegraphics[width=8cm,height=8cm]{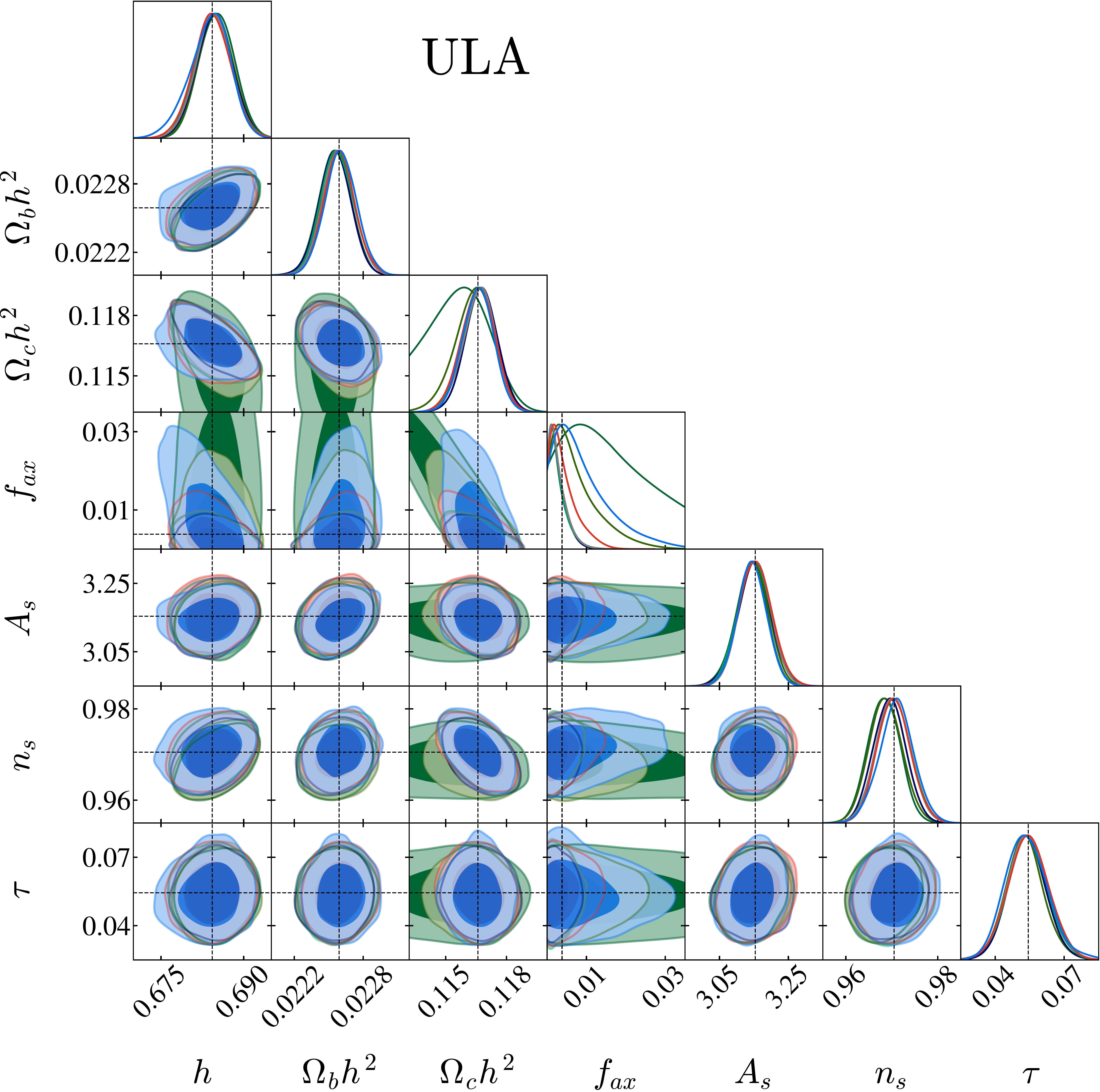}
\includegraphics[width=8cm,height=8cm]{6.pdf}
\caption{Reconstructed PDFs of cosmological parameters from Planck 2018 CMB and DESI-DR2 BAO data. Two models are considered in this work: the \texttt{ULA}-only model (left) and the \texttt{ULA+CPL} model (right). The red dashed lines in the right panel denote the $\Lambda$CDM predictions.} 
\label{allpdfs}
\end{figure*}

For the distance measurements, both the angle-averaged ($D_V/r_d$) and anisotropic ($D_H/r_d$, $D_M/r_d$) BAO measurements at seven redshifts are released by DESI~\citep{desi2}. Here, $D_{i\in V, H, M}$ and $r_d$ are comoving distances and sound horizon. Especially, to account for different correlations of the transverse and perpendicular BAO measurements, correlation coefficients are introduced in the covariance matrix for the likelihood function $\mathcal{L}_{\rm BAO}$ of the BAO measurements. Posterior distribution functions (PDFs) of the cosmological parameters are thus inferred from $\mathcal{P}(\theta|\hat d)\sim\mathcal{L}_{\rm CMB}(\hat d|\theta)\mathcal{L}_{\rm BAO}(\hat d|\theta)\pi(\theta)$ where $\pi(\theta)$ are parameter priors. 

\subsection{Validations with mock CMB and BAO datasets}

In this work, two scenarios have been tested. One is the $\Lambda$ cold dark matter ($\Lambda$CDM) model with ULA physics and is referred to as \texttt{ULA}. Another one is referred to as \texttt{ULA+CPL} which is a combination of the ULA model and the Chevallier-Polarski-Linder (CPL)~\citep{CPL_a, CPL_b} model described as $w(a)=w_0+w_a(1-a)$ where $w_0$ and $w_a$ are two free parameters and $a$ is the cosmological scale factor. 

We generated the mock datasets for the CMB power spectra and the BAO measurements assuming fiducial cosmological models with known ULA signals which are simulated with two ULA fractions at $f_{ax}=0.1$ and $f_{ax}=10^{-4}$ representing a non-zero ULA detection and a null detection scenarios, respectively. We assume a futuristic cosmic-variance-limited CMB survey with a fullsky coverage and noiseless observations and simulate the corresponding $C_{\ell}^{TT}$, $C_{\ell}^{EE}$ and $C_{\ell}^{TE}$ power spectra at the same $\ell$ range as the Planck data. We theoretically calculate the different distance ratios $D_V/r_d$, $D_M/r_d$ and $D_H/r_d$ at the seven redshifts where the DESI BAO measurements are made, while directly taking the covariance matrices and correlation coefficients from the DESI-DR2 distance measurements. Thus, there are four types of mock datasets used for the validations, i.e., \texttt{ULA} ($f_{ax}=0.1$), \texttt{ULA} ($f_{ax}=10^{-4}$), \texttt{ULA+CPL} ($f_{ax}=0.1$) and \texttt{ULA+CPL} ($f_{ax}=10^{-4}$). The test results in Fig. \ref{validation} demonstrate that the reconstructed PDFs of the cosmological parameters consistently peak at the fiducial values at the ULA mass $m_a=10^{-28}$ eV. Similar validations have been performed for other ULA masses and the reconstructed PDFs are also consistent with the theoretical models.

We also investigated the prior impact on the parameters. Specifically, we tested the inference of the ULA fraction $f_{ax}$ with a uniform prior on $\log_{10}(f_{ax})$. The inferred ULA fraction with such a prior is consistent with the input value if there is a nonzero ULA component. However, the inferred upper bound on the ULA fraction with a uniform prior on $\log_{10}(f_{ax})$ is slightly smaller than the one with a uniform prior on $f_{ax}$ directly if no ULA is assumed because the uniform prior on $\log_{10}(f_{ax})$ allows a dense population at extremely small value of $f_{ax}$. As a conservative approach, we adopt the uniform prior on the ULA fraction $f_{ax}$ as our benchmark test.

\section{Data analysis results and discussions}

We run the data analyses using the Planck 2018 CMB power spectra and DESI-DR2 BAO measurements and sample the posterior distributions from $\mathcal{P}(\theta|\hat d)\sim\mathcal{L}_{\rm CMB}(\hat d|\theta)\mathcal{L}_{\rm BAO}(\hat d|\theta)\pi(\theta)$ within the ULA mass range $10^{-31}<m_a<10^{-27}$ eV. As seen from Fig. \ref{fractionplot}, we achieve a 0.7\% upper bound on the ULA fraction at $10^{-28}$ eV. This is the most stringent constraint on the ULA fraction in the mass window within $10^{-31}<m_a<10^{-27}$ eV so far. The detailed PDFs for all the parameters are shown in Fig. \ref{allpdfs}.

While the inferred ULA fractions remain tightly constrained, we observe mild degeneracies between \( f_{ax} \) and cosmological parameters such as the spectral index \( n_s \) and amplitude \( A_s \) of the primordial curvature perturbation, the optical depth \( \tau \), and DM energy density fraction $\Omega_c h^2$. These arise because the ULA component can affect both the amplitudes and locations of the acoustic peaks of the CMB power spectra by either shifting the matter-radiation equality epoch when it is DM-like or accelerating the expansion history when it is DE-like. Moreover, the ISW effects constrained by the Planck low-$\ell$ data can further contribute to these correlations. However, the posterior distributions remain stable, indicating that the constraints are robust against these correlations.

Interestingly, the right panel of Fig. \ref{fractionplot} indicates that the current limit on the symmetry-breaking energy scale $f_a$ is roughly an order of magnitude higher than the energy scale of the grand unified theory (GUT) in the mass range $10^{-30}<m_a<10^{-28}$ eV. The joint constraints from future CMB and BAO may be able to approach the GUT scale and the ULA signatures can be a smoking gun of the GUT physics.

\begin{figure}
\includegraphics[width=8cm,height=8cm]{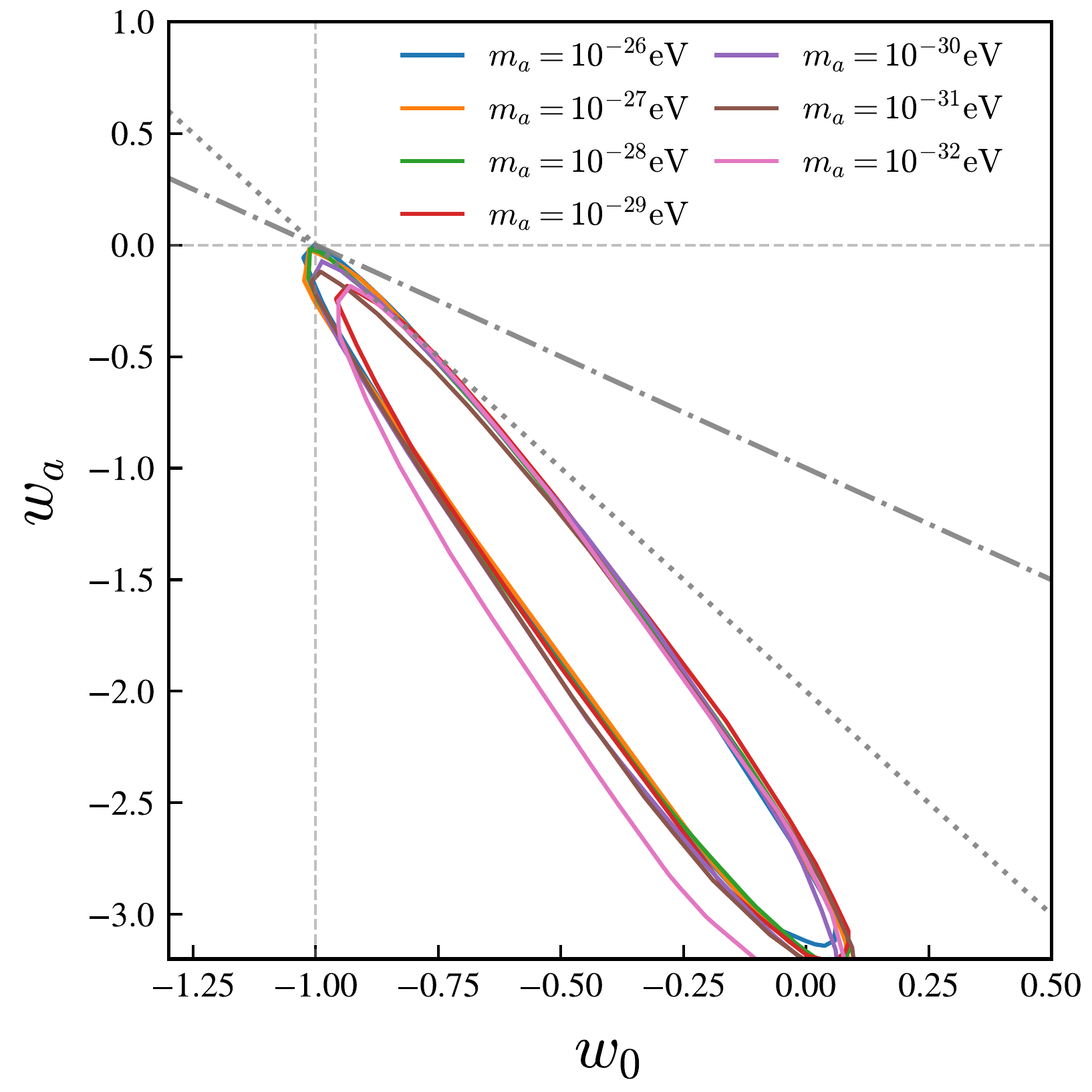}
\caption{Constraints on the $w_0$, $w_a$ parameters from the \texttt{ULA+CPL} model. The 95 \% confidence levels do not show as a strong tension as the model with \texttt{CPL}-only. The vertical and horizontal dashed lines represent the $\Lambda${\rm CDM} model. The tilted dot-dashed line corresponds to the constraint $1+w_0+w_a=0$ and the tilted dotted line is the constraint at $z=1$ when $w_0+(1-a)w_a=-1$, i.e., $1+w_0+w_a/2=0$.} 
\label{w0wa}
\end{figure}

\begin{figure}
\includegraphics[width=8cm,height=8cm]{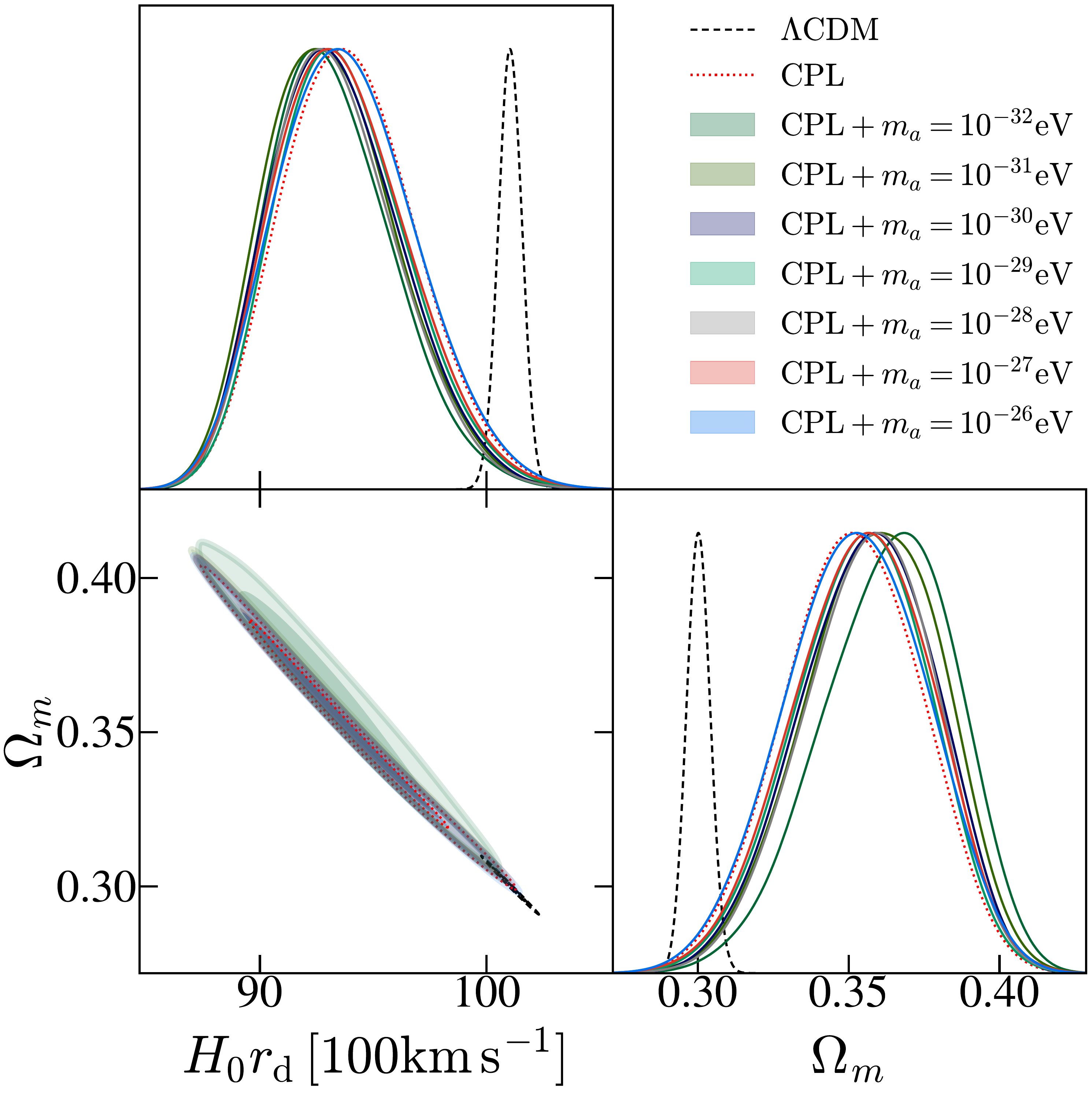}
\caption{Constraints on the $H_0r_d$ and $\Omega_m$ parameters from different models. The constraint from the \texttt{ULA}-only model is consistent with the $\Lambda$CDM model and the main difference is driven by the inclusion of the CPL model. } 
\label{rH}
\end{figure}

Fig. \ref{w0wa} shows that the constraints on the $w_0\mbox{-}w_a$ parameter space are slightly changed by including the ULA component which also has an evolving nature. This result implies that the ULA impact on the DDE maybe negligible given its ultra-low fraction. We note that this result is obtained with varying all nuisance parameters and if only an overall calibration parameter is assumed, the contours become consistent with the DESI result. Meanwhile, limited by the low fraction of $f_{ax}$, the ULAs do not change the constraints on the matter fraction and the sound horizon, as seen from Fig. \ref{rH} where the difference is mainly driven by the dominant CPL model.

\begin{figure}
\includegraphics[width=8cm,height=8cm]{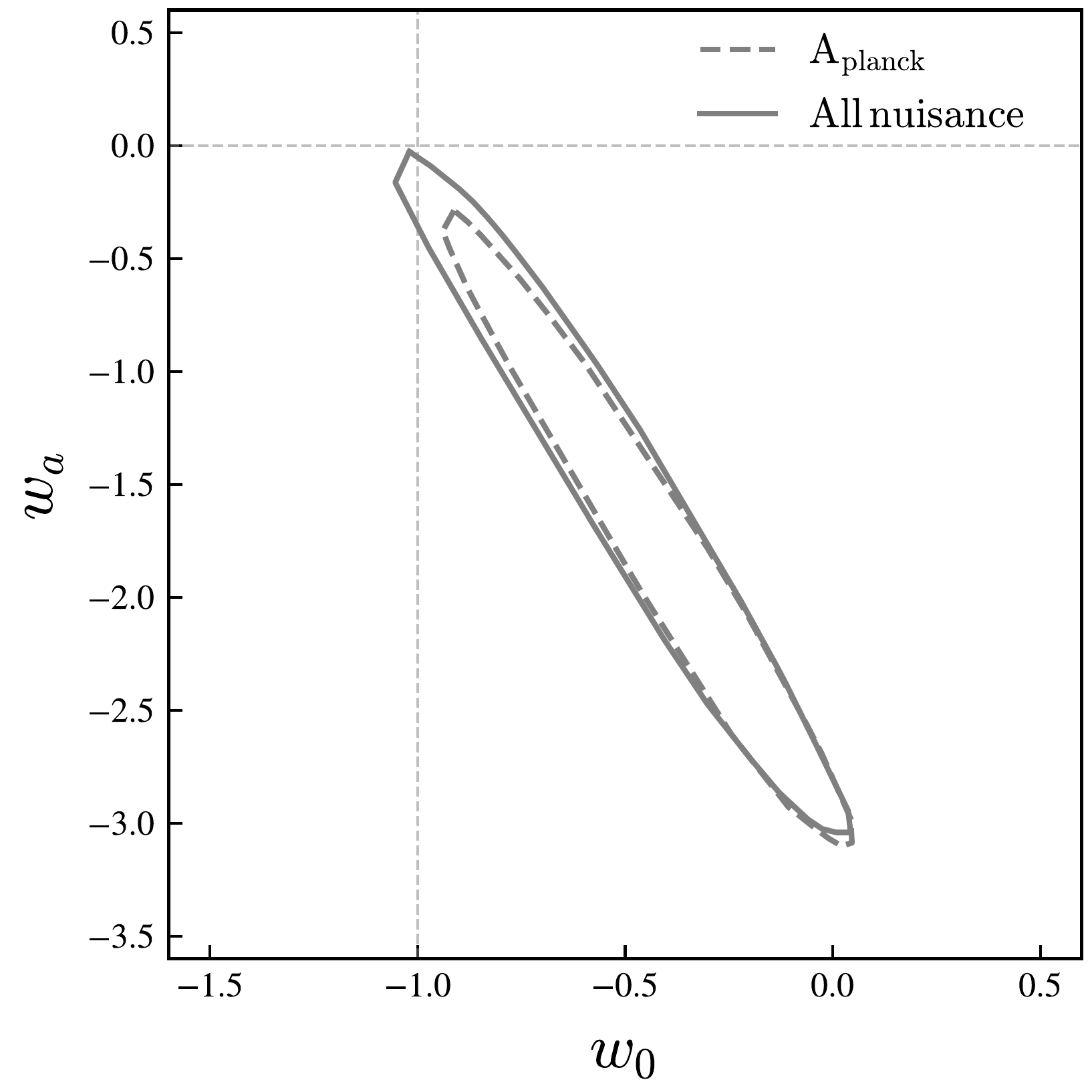}
\caption{The impact of nuisance parameters on the CPL model. We compare the inference of the CPL model under the assumptions of different nuisance parameters. The scenario with a single $A_{\rm Planck}$ parameter is consistent with the result in \cite{desi2}, while the deviation from the dark energy model with a cosmological constant is reduced as more nuisance parameters are included.} 
\label{comparison}
\end{figure}

The ULA is actually a quintessence field with an equation of state (EoS) $w$ satisfying $-1<w<0$, so it cannot resolve the issue with phantom crossing that requires $w<-1$~\citep{phantomcross} around $z\sim 0.4$~\citep{desi2}. This limitation is not just technical but fundamental since the ULAs are confined to quintessence-like behavior as minimally coupled scalar fields with canonical kinetic terms, and cannot cross the so-called phantom divide~\citep{phantomcross}. 
Our results indicate that including ULAs can help reduce the statistical tension between $\Lambda$CDM and DDE models. This reduction should be viewed primarily as an effective degeneracy rather than a fundamental physical resolution, and its precise impact depends on the mass of the ULA considered. When marginalising over axion masses, the overall effect is a net improvement, though individual fixed-mass cases may show varying levels of consistency with $\Lambda$CDM. The degree of mitigation ultimately depends on the precision of the data and the statistical significance of any deviation from $w=-1$.
Should future surveys confirm phantom-like evolution with increasing confidence, models beyond single scalar fields, such as non-canonical fields~\citep{kessence}, non-minimal couplings~\citep{nonminimal2} and modified gravity theories~\citep{nonminimal}, would be necessary to fully explain the observations.

In addition, in Sec. \ref{Method}, we only consider the potential with $n=1$ but do not study the models with $n>1$~\citep{axiclass}. The models with $n>1$ behave like early dark energy (EDE) which mainly affects the radiation-dominated epoch and behaves like radiation in the late universe. A recent analysis has shown that the inclusion of the EDE scenario can also reduce the preference for the dynamical dark energy and alleviate the Hubble tension~\citep{qghEDE}.

The CMB power spectra released by Planck are derived from the foreground separated maps and contain different foreground residuals which may not be negligible to the ULA-induced CMB signatures. Specifically, these residuals consist of cosmic infrared background, thermal and kinetic Sunyaev-Zel'dovich (tSZ/kSZ) effects, residual Galactic foreground polarization, as well as instrumental effects such as calibrations. 

We assigned different parameters to these residuals and marginalized all of these nuisance parameters for the inference of the cosmological parameters. These nuisance parameters included in the \texttt{Plik} package~\citep{pk2018} are associated with either uniform or Gaussian priors, which are also included in the Bayesian analysis discussed in Sec. \ref{Method}.

To validate the CMB and BAO likelihood functions, we ran an analysis with the $\Lambda${\rm CDM}+\texttt{CPL} model and with all the nuisance parameters included. The inferred parameters are consistent with the ones using \texttt{cobaya}~\citep{cobaya}.

We also ran the analysis with Planck likelihood package \texttt{Plik\_lite} which only assumes a Gaussian prior on a single polarization calibration factor $A_{\rm Planck}$ while fixing other nuisance parameters as implemented in~\cite{rogersetal2023}, our result is the same as the one in~\cite{desi2}, as shown in Fig. \ref{comparison}.

\section{Conclusions}

By jointly analyzing the Planck 2018 CMB data and seven BAO measurements from DESI data release 2, we obtain an improved constraints on the ULA fractions in the mass range $10^{-30}{\rm eV}<m_a<10^{-28}{\rm eV}$. Also, the inferred $w_0\mbox{-}w_a$ parameters show negligible impact from the ULA component under different assumptions for nuisance parameters.

The next-generation CMB polarization experiments will obtain high sensitivity datasets~\citep{so, pico, litebird}, and more precise and independent BAO measurements will also be obtained at both optical~\citep{lsst} and radio wavelengths~\citep{ska}. The enormous constraining power of the future joint datasets may shed light on the nature of dark energy and find evidence of the ULA signatures.

All of these constraints presented in this work are obtained from the gravitational effects of the ULAs which may also leave interesting signatures through other mechanisms. The ULAs may be coupled to the electromagnetic field and generate a cosmological rotation effect~\citep{cb1,cb2} that can be precisely measured by the next-generation CMB polarization experiments. In addition, the ULAs can power the isocurvature perturbations~\citep{isocurvature} and the CMB polarization may place new limits on the ULA physics. Moreover, high precision matter clustering measurements~\citep{lsst} at different wavelengths can significantly break the degeneracies among cosmological parameters so the ULA signatures can be better constrained by a multitracer dataset in the future.

\begin{acknowledgments}
This work is supported by the starting grant of USTC. F. B. A. acknowledges the support of the CAS. We acknowledge the use of the \axioncamb~\citep{renee2015}, \emcee~\citep{emcee} packages.
\end{acknowledgments}

\bibliography{axionnew}

\end{document}